# Colored visible light metamaterials based on random dendritic cells

K.Song, H.L.Ma, B.Q.Liu, X.P.Zhao*

*Smart Materials Laboratory, Department of Applied Physics, Northwestern Polytechnical University, Xi'an 710129, P R China    E-mail: xpzhao@nwpu.edu.cn*

Optical metamaterials (OMs) at visible wavelengths have been extensively developed. OMs reported recently are all composed of periodic structure, and fabricated by top-down approaches. Here demonstrated are the colored visible light frequency metamaterials composed of double layer array disordered and geometrical variational dendritic cells as well as fabricated by a novel bottom-up approach. The experiment demonstrates that the OMs composed of random silver dendritic cells which caused the appearance of multiple transmission passbands at red and yellow light frequencies. Moreover, the slab focusing experiment reveals a clear point image in the range of half-wavelength with an intensity of 5% higher than that of the light source. The colored OMs proposed will create a new way to prepare the cloak and the perfect lens suitable for optical frequency.

Left-handed metamaterials (LHMs) at optical wavelengths, with simultaneously negative dielectric permittivity $\varepsilon$ and magnetic permeability $\mu$ [1-5], have been developed as research focus in this subject [6,7]. Similar to photonic crystals and other artificial materials with periodical structure, the LHMs reported recently are all composed of periodic structure. Since the dimension of unit cell is much smaller than the corresponding wavelength, the fabrication of LHMs at the optical frequency has become a very difficult problem.

Based on the theory of periodically structured LHMs, the fabrication of LHM consistently employs the top-down approach of e-beam lithography or focused ion beam technique, such as arrays of U-shaped Au nanostructures [8,9], and double-fishnet structures [10]. Valentine et al. have experimentally demonstrated the first 3D fish-net negative refractive index metamaterial at infrared wavelengths [11]. Recently, García-Meca et al. have further realized negative refractive index at visible wavelengths by multilayered fish-net metamaterial with higher transmission and lower loss [12]. Gorkunov et al. have demonstrated that even a weak microscopic disorder may lead to a substantial modification of the metamaterial magnetic properties. Moreover, a 10% deviation in the parameters of the microscopic resonant elements may lead to a substantial suppression of the wave propagation in a wide frequency range [13]. Golluba et al have presented both the theoretic and experimental results of geometrical variations in generalization [14]. Zhou et al. proposed disorder effects of LHMs with unitary dendritic structure cell [15]. However, all these researches have achieved disordering array and geometrical variations separately, and what's more, they are at microwave and infrared wavelengths [16,17]. Recently, Jen et al. have realized negative refractive index and permeability in the visible regime by means of a disordered silver metamaterial thin film [18].

Here, we proposed a colored optical metamaterials (OMs) at visible wavelength based on random dendritic cells, aiming to disorder array and geometrical variations at the same time. The first layer of random silver dendritic cells was prepared on Indium Tin Oxide (ITO) conductive glass substrate by electrochemical deposition, and then the dielectric medium of polyvinyl alcohol (PVA) film was located on the first layer of silver dendritic cells. Finally, the sandwich-like structure was fabricated by depositing the second layer silver dendritic cells on the PVA film, while the flat focusing experiment was used to demonstrate the properties of the samples prepared beforehand.

Fig. 1(a) shows the model of the OMs operating at visible frequency based on random dendritic cells. The cycle of dendritic cells is array disordered while the geometrical structure is variational. a and b represent the lengths of the first and second offshoots of the dendritic cells respectively, while $w_a$ and $w_b$ refer to the width.

Based on the fact that the electromagnetic properties of LHMs are basically an effect of micro or nano inclusions in contrast to the photonic crystal technology whose properties are dependent on array effects, while the dendritic cell sizes are controlled round two given structures statistically. (Table.1).

Tab. 1  The dimension parameters of the two structures

|       | Structure 1 | Structure 2 |
|-------|-------------|-------------|
| a     | 135         | 120         |
| $w_a$ | 75          | 60          |
| b     | 50          | 25          |
| $w_b$ | 40          | 20          |

the thickness of Ag dendritic cell is 35nm

the thickness of PVA film is 20nm

For incident plane waves with the magnetic field parallel to the dendrite cells, a strong magnetic resonance may be obtained from the two dendrite cells face to face at a certain frequency, which are mainly produced by the width "$w_a$" of the dendritic cells, perpendicular to $H$ direction. Meanwhile, the negative permittivity is obtained by the dipolar resonance arising from the length "a" of the dendritic cells in $E$ direction.

By employing the method in ref. [14], we have calculated the effect of the geometrical variations at visible frequencies, which shows that the resonance behavior of two given structures could still exist should the cell variation remain below ±7%.

According to the simulated effective parameters of structures 1 and 2, the OMs composed of random silver dendritic cells, dielectric medium PVA and random silver dendritic cells was fabricated as illustrated in Fig.1(b) [19].

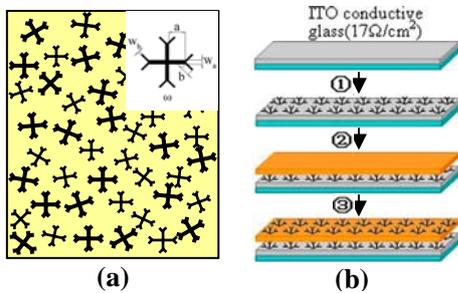

**Fig. 1** (a) Schematic illustration of the silver dendritic cells and (b) the preparation of the sandwich-like structure: ① Electrochemical deposition of the first layer of silver dendritic cells, ② Coating with PVA film, ③ Electrochemical deposition of the second layer of silver dendritic cells.

Firstly, the silver dendritic cells were prepared by nano assembling approach of electrochemical deposition using the setup illustrated in Fig. 1(b). In this procedure, a conductive glass (50mm×10mm) with ITO film in thickness of 100nm and a planar and smooth silver slice (55mm×12mm) were used as cathode and anode respectively, and the conductive glass substrates were cleaned by procedures previously reported [20]. Two PVC slices in thickness of 0.6mm were employed to control the space between the cathode and the anode for further controlling the size and the density of the silver dendritic cells prepared so as to be consistent with the parameters of the model proposed. A complicated electrolyte was used to control the length and the width of the branches of the silver dendritic cells prepared to be consistent with the parameters of structures 1 and 2, and was prepared as follows: 1.2g PEG-20000 was dissolved in 5ml ultra pure water under magnetic stirring before 5ml AgNO$_3$ aqueous solution (0.2mg/ml) was added. Meanwhile, the solution was unceasingly stirred and irradiated by tungsten lamp until its color became light pink. The electrolyte was inhaled into the space between the cathode and the anode by capillary force, and the electrochemical deposition was carried out at DC voltage of 0.9V for 90s.

Secondly, the PVA film in a thickness of 20nm~50nm, as an electrical-insulated interlayer in the OMs, was prepared and located on the first layer silver dendritic cells by spin coating method at a speed of 1000 rpm with PVA aqueous solution (0.5wt %). Later, the sample was put in a vacuum oven and the PVA film was heated for 0.5 hours at 170℃, preventing it from dissolving in water.

Finally, the second layer of silver dendritic cells were deposited on the PVA film with the first layer of silver dendritic cells as cathode, and the electrochemical deposition was carried out at DC voltage of 2.9V for 240s.

The OMs composed of random silver dendritic cells, dielectric medium PVA and random silver dendritic cells was prepared. The sandwich structural sample is highly transparent with its area being of 10×20(mm)$^2$. Specifically, this sample can be prepared in flexible size of area on demand.

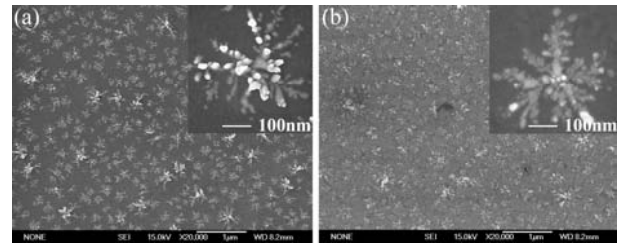

**FIG. 2** (a) SEM images of the silver dendrtic cells on ITO film, and the insert is a high magnification image of a unit cell. (b) SEM images of the silver dendrtic cells on PVA film.

Fig 2(a) shows the SEM image of the silver dendrites on ITO film. It is observed that the silver dendritic cells in diameter of about 150nm~500nm arrayed out of order on its substrate. It is composed of main trunk and side branches with length of 50nm~100nm and 15nm~60nm respectively, and the width of the branch is about 20nm~65nm. The silver dendritic cell prepared on PVA film is shown in Fig 2(b), with almost the same size as that prepared on ITO film. Additionally, it is also found that the size of the silver dendritic cells prepared beforehand is approximately consistent with the dimension parameters of the model proposed.

The visible transmission spectra of the silver dendritic array was recorded at room temperature by a UV-4100 spectrophotometer in the wavelength range of 400-800nm. The visible light focusing experiment was performed on a home-built visible light focusing apparatus, with the monochromatic light produced by the monochromator (Omni-λ300, Zolix Instruments Co., Ltd.) using a xenon lamp as the light source (LHX150, Spectrum: 200-1800 nm, Zolix Instruments Co., Ltd.). The intensity of transmitted light was measured by a setup consisting of a fiber optic spectrometer (USB2000, Ocean Optics, Inc.) and a fiber probe as the receiver, and the divergent beam was produced by adjusting the locations of the convex lens and the diaphragm, while the diameter of the light spot was regulated to about 0.3mm by adjusting the diaphragm. Then the sample was fixed on an underprop, and the fiber probe was moved near the sample. At the same time, the diaphragm of the monochromator was adjusted to make the intensity of transmitted light at about 12000 counts. The distance between the fiber probe and LHM lens was controlled by the one-axis fine stages controller (Travel Accuracy: 1.0 nm, Piezo Positioning System FINE-01) during the experiments, and the intensity of transmitted light along the *X*-axis was measured by fiber probe. During experiments, the steps are 20 nm when the distance between the fiber probe and the sample is 0-800 nm.

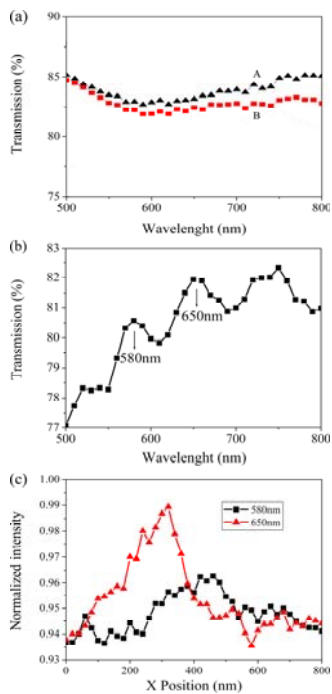

**FIG. 3** (a) Visible light transmission spectrum of ITO conductive glass (curve A) and ITO conductive glass coated with PVA film (curve B), (b) Visible light transmission spectrum of the dendritic metamaterials, which reveals two pass-band peaks at wavelength of 580 nm and 650 nm, respectively. (c) Flat lens focusing curves in X position, the yellow and red light transmitted through the sample were focused respectively.

Fig.3(a),(b) show the visible light transmission spectrum of ITO conductive glass (curve A), ITO conductive glass coated with PVA film (curve B) and the sandwich structure, respectively. It can be observed that neither the ITO conductive glass nor the ITO conductive glass coated with PVA film present pass-band peaks when being tested individually. However, the sandwich structure resulted in the appearance of two pass-band peaks at wavelength of 650nm and 580nm respectively, which means the response of incident light is produced by the sandwich OMs sample. For further demonstration of our results, the slab focusing experiment was performed. Fig. 3(c) shows the focusing results of the sample prepared beforehand along X direction, indicating evidently that the visible light in wavelength of 580nm and 650nm transmitted through the sample and focused at the point of 420nm and 320nm away(both in the half wavelength region) from the test sample respectively. Moreover, the red curve reveals a focused peak with an intensity of 5% higher than that of the light source at 650nm wavelength.

In conclusion, based on the fact that the electromagnetic properties of LHMs are resulted from micro or nano inclusions in contrast to the photonic crystal relying on array effects, we propose the colored OMs at visible frequency based on double layer silver random dendritic cells, which is in disorder array and geometrical variations. The calculation shows that the whole left-handed resonance behavior could still exist should the geometrical variations remain below ±7%. The sandwich structural sample was fabricated with random silver dendritic cells, PVA and random silver dendritic cells, and the sample prepared beforehand showed colored passbands at yellow light 580 nm and red light 650 nm. At the same time, the slab focusing experiment reveals a clearly focused point image in 3λ/4 and λ/2 region with an intensity of 2.5% and 5% higher than that of light source at 580 nm and 650 nm wavelength, respectively. In addition, the samples were with a large area of 10mm×20mm, which greatly exceeded the samples prepared by top-down approach (in scale of square micrometer). In a nutshell, the colored OMs composed of random cells will create a new way to prepare the cloak and the perfect lens suitable for optical wavelengths.

This work was supported by the National Nature Science Foundation of China under Grant No.50872113, 50936002.